\documentclass[a4paper,fleqn,usenatbib,useAMS]{mnras}

\usepackage{graphicx}	
\usepackage{amsmath}	
\usepackage{amssymb}	
\usepackage{multicol}        
\usepackage{bm}		
\usepackage{pdflscape}	
\usepackage{hyperref}
\usepackage{xcolor}

\usepackage[T1]{fontenc}
\usepackage{ae,aecompl}

\usepackage{newtxtext,newtxmath}

\title[Kozai inside RMMRs]{Kozai-Lidov Mechanism inside Retrograde Mean Motion Resonances}

\author[Yukun Huang et al.]{
Yukun Huang,$^{1}$\thanks{E-mail: puluobi@gmail.com}
Miao Li,$^{1}$\thanks{E-mail: limiaotsinghua14@163.com}
Junfeng Li$^{1}$\thanks{E-mail: lijunf@mail.tsinghua.edu.cn}
and Shengping Gong$^{1}$\thanks{E-mail: gongsp@tsinghua.edu.cn}
\\
$^{1}$School of Aerospace Engineering, Tsinghua University Beijing, China, 100086\\
}

\date{Received 2018 July 15.\@ Revised 2018 August 15.\@ Accepted 2018 September 13.}

\pubyear{2018}

\begin{document}
\label{firstpage}
\pagerange{\pageref{firstpage}--\pageref{lastpage}}
\maketitle

\begin{abstract}
	As the discoveries of more minor bodies in retrograde resonances with giant planets, such as 2015 BZ509 and 2006 RJ2, our curiosity about the Kozai-Lidov dynamics inside the retrograde resonance has been sparked. In this study, we focus on the 3D retrograde resonance problem and investigate how the resonant dynamics of a minor body impacts on its own Kozai-Lidov cycle. Firstly we deduce the action-angle variables and canonical transformations that deal with the retrograde orbit specifically. After obtaining the dominant Hamiltonian of this problem, we then carry out the numerical averaging process in closed form to generate phase-space portraits on a $e-\omega$ space. The retrograde 1:1 resonance is particularly scrutinized in detail, and numerical results from a CRTBP model shows a great agreement with the our semi-analytical portraits. On this basis, we inspect two real minor bodies currently trapped in retrograde 1:1 mean motion resonance. It is shown that they have different Kozai-Lidov states, which can be used to analyze the stability of their unique resonances. In the end, we further inspect the Kozai-Lidov dynamics inside the 2:1 and 2:5 retrograde resonance, and find distinct dynamical bifurcations of equilibrium points on phase-space portraits.
\end{abstract}




\section{Introduction}\label{sec:1}

The Kozai-Lidov mechanism (or Lidov-Kozai mechanism) of small bodies in the solar system has been studied for years. It was first noticed by \citet{Lidov:1962du} when analyzing the evolution of orbits of artificial satellites of Earth. Meanwhile, \citet{Kozai:1962fa} realized that the secular perturbations of asteroids with high inclination and eccentricity in a Sun-Jupiter system should have great similarities with those of the satellites problem.

Nowadays, the Kozai-Lidov mechanism is considered to be one of the most fundamental dynamical process that influences the basic structure of both the solar system and the extrasolar system \citep{Shevchenko:2016uj}. During the early 1990s, \citet{Bailey:1992vj} used a semi-analytical model that treats the Jupiter as a homogeneous ring to study the origins of sungrazing comets. On this basis, \citet{Thomas:1996in} extended this idea to including the gravitational perturbations of all the four giant planets, and carried out a systematic study of the Kozai-Lidov dynamics in the outer Solar System (i.e.\ semi-major axis between $40$ and $100$ au).
They showed that the librational islands around $\omega = \pm 90^\circ$ are more stable than those around $\omega = 0^\circ$ and $180^\circ$ when the eccentricity is extremely high.
Moreover, \citet{Michel:1996us} followed an analogous method that takes into account all the terrestrial and Jovian planets to study the Kozai-Lidov mechanism in the neighbourhood of the Earth (i.e.\ $a < 2$ au). Their research demonstrated that the librations of $\omega$ around $0^\circ$ and $180^\circ$ can actually occur at very low inclinations, as long as the semi-major axis of the asteroid is close to that of the main perturber. Similar to this, the Kozai-Lidov dynamics for minor bodies in a large range of semi-major axis (i.e.\ $a$ from 1 au to 18 au) was studied in detail by \citet{Gronchi:wj}. In addition, they presented a theorem that a stable Kozai-Lidov state which prevents the orbits of minor bodies from node crossings always exists for every value of $a$ and any number of perturbing planets.

The pure Kozai-Lidov mechanism in the solar system can be well understood by a semi-analytical averaging model. However, if the mean motion resonance is taken into account, the Kozai-Lidov dynamics becomes much more different and hard to fathom. As \citet{Morbidelli:2002tn} pointed out in his book, the study of the secular dynamics in regular regions of mean motion resonances is one of the most complicated issues in celestial mechanics. And of course, the Kozai-Lidov dynamics inside a mean motion resonance is not an exception. The study of this problem is getting more attention in recent years.

After \citet{Kozai:1962fa} published his analytical theory on the pure secular perturbations of asteroids, he presented another semi-analytical method in \citet{Kozai:1985ut} that handles the secular perturbations of asteroids inside some mean motion resonance. It was shown with the Pluto-Neptune system that his method agrees well with the numerical integration results. Additionally, it was also found that for some cases of resonances, the argument of pericenter $\omega$ can librate around angles other than $k90^\circ$, which was called the asymmetric libration afterwards. \citet{Morbidelli:1993ul} elucidated the Kozai-Lidov dynamics inside the 2:1 and 3:2 mean motion commensurabilities with the Hamiltonian canonical transformation and averaging over the fast angles of the planet. A similar study was also accomplished by \cite{Morbidelli:1995jf} to understand the resonant and secular dynamics of the Kuiper belt objects. \citet{Nesvorny:2002vz} analyzed the Kozai-Lidov cycles inside a prograde 1:1 co-orbital resonance. Unlike previous studies, they considered the eccentricity and inclination of both the perturbed and the perturbing body and found new types of orbits.

Besides shaping the dynamical structure of the inner solar system, The Kozai-Lidov mechanism inside the mean motion resonance significantly influences in the Kuiper belt as well. \citet{Fernandez:2004kh} showed in their research that Kozai-Lidov dynamics may prompt the perihelion distance of scattered disk objects to a large value, which corroborates the idea that they may originate from the Oort cloud. Furthermore, \citet{Gomes:2005fl} put forth the idea that the combined effects of the high-order MMR with Neptune along with the Kozai-Lidov mechanism supply a transfer process from the scattered disk to the high-perihelion scattered disk. Moreover, \citet{Gallardo:2006gd} calculated the critical inclination of the Kozai-Lidov dynamics near the high-order resonance analytically with the series expansion method. Another study about the resonant Kozai-Lidov effect in the Kuiper Belt was made by \citet{Wan:2007ed}. They adopted an analytical method based on series expansion to the cases of 1:2 and 2:3 mean motion resonances. In addition, \citet{Gallardo:2012gs} focused on the Kozai-Lidov dynamics of the resonant and nonresonant TNOs. They noted that the present distribution of orbital inclinations of the TNOs can not be explained by Kozai-Lidov dynamics with Neptune, no matter resonant or nonresonant, as the low inclination of the primordial population does not allow it to generate a more diverse distribution of orbital inclinations and eccentricities of TNOs.

Recent surveys showed that the clustering of the argument of pericenter $\omega$ around $0^\circ$ of distant Kuiper Belt objects can be possibly explained by the Kozai-Lidov mechanism with an undiscovered planet \citep{Trujillo:2014iha}. \citet{2016AJ....151...22B} put forward this idea that a distant giant planet with $10 m_{\earth}$, which is called Planet Nine these days, may lead to the clusterings of $\Delta\omega$, $\Delta\Omega$, and $\Delta\varpi$ of distant KBOs. In addition, high-inclination scattered disk bodies can be generated by Planet Nine through secular gravitational effect \citep{2016ApJ...833L...3B}.

In summary, it is widely acknowledged over the past decades that Kozai-Lidov dynamics plays a key role in the dynamical evolution of the solar system, and it may also shed light upon the origins of mysterious minor bodies in the inner and outer solar system.

Recently, the newly discovered asteroid 2015 BZ509, drew a lot of attention to its unique resonant dynamics. It is the first confirmed asteroid in a retrograde 1:1 resonance (or co-orbital resonance) with Jupiter and has a long-term stability for around 1 Myr \citep{Wiegert:2017fj,Morais:2017fn}. It was shown recently that it may probably be captured from the interstellar space \citep{Namouni:2018hl} with massive numerical simulations. And \citet{2017arXiv170400550N} also demonstrated that 2015 BZ509 lies near the peak of co-orbital capture efficiency. Before 2015 BZ509 was found, several studies about the retrograde mean motion resonance had already been carried out numerically and analytically \citep{2013CeMDA.117..405M,Morais:2013ka,2012MNRAS.424...52M}. Particularly, \citet{Morais:2016kq} studied the stability with chaos indicator near the co-orbital region in the three-dimensional (3D) space. They considered both the resonant and the Kozai-Lidov effect, but found no simultaneous small-amplitude librations of both the resonant angle $\varphi$ and the argument of pericenter $\omega$. We will show later that their numerical results get well understood in our paper.

However, although several studies about the retrograde resonances have been carried out these days, no special attention was paid to the dynamical structure of the Kozai-Lidov mechanism inside the resonances. So in this paper, we focus specifically on the Kozai-Lidov mechanism inside a retrograde resonance, or in other words, the resonant Kozai-Lidov mechanism for retrograde orbits. It is noteworthy that our study is carried out under the frame of the circular restricted three body problem (CRTBP) model, that is, we ignore the planetary eccentricity and inclination, as well as its slow precession. And it is based on our previous work on the dynamics of the retrograde mean motion resonance \citep{Huang:2018ey}, in which a planar retrograde resonance can be well-understood with the phase-space portraits generated by a semi-analytical method \citet{Morbidelli:2002tn}. On this basis, we re-consider the asteroidal inclination and study the retrograde resonance in 3D space in this work.

Before we start our paper, there is one trivial but essential thing needs to be clarified, that is, if a retrograde minor body is outside a retrograde mean motion resonance, then its Kozai-Lidov mechanism is identical to its prograde counterpart (i.e. the orbit whose $I_\textrm{pro} = 180^\circ - I_\textrm{retro}$). It is easy to understand since in the assumption of Kozai-Lidov dynamics, the planet can be replaced by a homogeneous ring. And in the `eyes' of a ring, there is no difference at all between a prograde orbit and a retrograde orbit with the same relative inclination. The symmetry of this problem ensures they have the same secular dynamical evolution. However, the resonant Kozai-Lidov dynamics of a prograde orbit and a retrograde orbit are completely divergent, as they have their respective definitions of critical angles. Therefore, different restrictions of resonant angles shape the Kozai-Lidov dynamics in totally different ways. In addition, we will see the difference between prograde and retrograde resonant dynamics later in our work.

Finally, our work is organized as follows: In Section~\ref{sec:2}, we adapt the semi-analytical method described in \citet{Morbidelli:2002tn} and \citet{Moons:1993fv} to the case of the retrograde mean motion resonance, and deduce similar Hamiltonian canonical transformations that tackle the 3D resonance problem. A nice and simple one d.o.f model is presented in the end of this section, which can be adopted into resonances with any ratio. Then, in Section~\ref{sec:3}, we focus on the particular retrograde 1:1 resonance and analyze its dynamical structure of the Kozai-Lidov cycles of angle $\omega$ in 3D space. We numerical calculate the averaged Hamiltonian on an $e-\omega$ grid and then plots its corresponding phase-space portrait, taking the influence of the amplitude of the resonant angle into account. Moreover, we compare the numerical results in a CRTBP model with the semi-analytical phase-space portraits. In Section~\ref{sec:4}, we utilize this method to analyze the Kozai-Lidov state of two real minor bodies inside the retrograde 1:1 resonance, they are, 2015 BZ509 and 2006 RJ2, the latter is confirmed to be in the retrograde 1:1 resonance with Saturn by \citet{Li:2018kn}. Finally, in Section~\ref{sec:5}, we apply this methodology to two more retrograde resonances, 2:1 and 2:5, and find different bifurcations of equilibrium points with phase-space portraits.

\section{Semi-analytical Model}\label{sec:2}

In our last paper \citet{Huang:2018ey}, we fully investigate the dynamic portrait of the planar retrograde 1:1 mean motion resonance. Next, we briefly summarize the canonical action-angle variables used for the retrograde orbit and the retrograde resonance, and then present the Hamiltonian transformation to study the Kozai-Lidov dynamics inside the resonance. After we get an one d.o.f integrable model for the problem, we can numerically evaluate the Hamiltonian to understand the complicated dynamics intuitively with the straightforward phase-space portraits. Such a method is often called the semi-analytical method, as it combines the process for analytically reducing the system to a single angle variable, and numerically calculating its dominant Hamiltonian on a parameter space.

\subsection{Action-angle Variables for Retrograde MMRs}
We start from the following retrograde canonical Poincar{\'e} action-angle variables for a test particle \citep{Huang:2018ey, Gayon:2009ct, Namouni:2014dz}:
\begin{equation}\label{eq:retrograde_poincare_variables}
	\begin{aligned}
		\Lambda & = L,\quad     & \lambda & = M + \omega - \Omega,         \\
		P       & = L - G,\quad & p       & = -\left(\omega-\Omega\right), \\
		Q       & = G + H,\quad & q       & = \Omega,                      \\
	\end{aligned}
\end{equation}

where $M,\omega,\Omega$ are mean anomaly, argument of perihelion and longitude of node, and $L = \sqrt{a},\ G = L\sqrt{1-e^2},\ H = G\cos{i}$ are the corresponding conjugate actions. It is important to keep in mind that the mean longitude $\lambda$ for retrograde orbit is now $M + \omega - \Omega$, rather than the sum of these three angles in the prograde case. Again, the longitude of perihelion $\varpi$ is now defined as $\omega - \Omega$ for the retrograde orbit, rather than $\omega + \Omega$.

To investigate a particular retrograde resonant system, it is convenient to introduce the following action-angle variables for a retrograde $k^\prime : k$ resonance:
\begin{equation}\label{eq:retrograde_canonical_variables}
	\begin{aligned}
		S   & = P,                                    & \sigma   & = \cfrac{k \lambda- k^\prime \lambda^{\prime}+\left(k+k^\prime\right)p}{\left(k+k^\prime\right)}, \\
		S_z & = Q,                                    & \sigma_z & = \cfrac{k \lambda- k^\prime \lambda^{\prime}+\left(k+k^\prime\right)q}{\left(k+k^\prime\right)}, \\
		N   & =-\cfrac{k+k^\prime}{k} \Lambda + P +Q, & \nu      & = -\cfrac{k \lambda- k^\prime \lambda^{\prime}}{\left(k+k^\prime\right)},
	\end{aligned}
\end{equation}
where $\sigma$ and $\sigma_z$ are the critical angles, while $S$ and $P$ are the corresponding conjugate actions. The action $N$ is a first integral of the retrograde system, and it can be simplified as

\begin{equation}\label{eq:N_def}
	N = \sqrt{a} \left( \sqrt{(1-e^2)} \cos{i} -\cfrac{k^\prime}{k}\right).
\end{equation}

If we restrict the motion of the test particle to the plane, which corresponds to the planar circular restricted three body problem, the Hamiltonian depends solely on one resonant angle, $\sigma$. Then we get a one d.o.f integrable approximation of the retrograde MMR. In \citet{Huang:2018ey}, we have developed a systematic methodology to analyze any retrograde MMRs in closed form with semi-analytical tool. From now on, we put our interests into the more complicated three-dimensional (3D) case.

\subsection{Kozai-Lidov Dynamics Inside Retrograde MMRs}
Firstly, let's assume a test particle is in a retrograde $k^\prime : k$ resonance with a planet. If the inclination is not excessively large and the eccentricity of the test particle is significantly larger than that of the planet, we can define an integrable approximation $\mathcal{H}_{\sigma}$ with the following form \citep{Morbidelli:2002tn}:
\begin{equation}\label{eq:H_sigma}
	\mathcal{H}_{\sigma} = \mathcal{H}_{0} (N,S,S_z) + \mathcal{F}_{\sigma} (N,S,S_z,\sigma),
\end{equation}
where
\begin{equation}\label{eq:H_0}
	\mathcal{H}_0 = -\cfrac{{\left(k^\prime+k\right)}^2}{2k^2{\left(N-S-S_z\right)}^2} + n^\prime \cfrac{k^\prime}{\left(k^\prime+k\right)}\left(N-S-Sz\right)
\end{equation}
is the Kepler term, and $\mathcal{F}_{\sigma}$ is the perturbation that depends only on angle $\sigma$. Because $\mathcal{H}_{\sigma}$ is integrable, we introduce the following Arnold action-angle variables:
\begin{equation}\label{eq:arnold_variables}
	\begin{aligned}
		\psi     & = \frac{2\pi}{T_\sigma}t, \quad              & J     & = \frac{1}{2\pi} \oint S \mathrm{d} \sigma, \\
		\psi_z   & = \sigma_z - \rho_z(\psi,J,J_z,J_\nu), \quad & J_z   & = S_z,                                      \\
		\psi_\nu & = \nu - \rho_\nu(\psi,J,J_z,J_\nu), \quad    & J_\nu & = N,                                        \\
	\end{aligned}
\end{equation}
where $T_\sigma$ denotes the period of $\sigma$, while $\rho_z$ and $\rho_\nu$ are also periodic with zero average. The geometric meaning of the action $J$ is obvious. It denotes the area bounded by the librating trajectory in the $(S,\sigma)$ plane.
With the new set of action-angle variables, $\mathcal{H}_{\sigma}$ now depends only on the action variables, that is
\begin{equation}\label{eq:H_sigma_arnold}
	\mathcal{H}_{\sigma} = \mathcal{F}_{0} (J,J_z,J_\nu).
\end{equation}
In order to investigate the Kozai-Lidov mechanism inside a retrograde MMR, we need to introduce the part of Hamiltonian that depends not only on critical angle $\sigma$. So in the new variables, the complete Hamiltonian describing the Kozai-Lidov dynamics reads~\footnote{The angle $\psi_\nu$ does not show up in $\mathcal{H}_{1}$ because we do not take into account the planetary inclination and eccentricity.}
\begin{equation}\label{eq:H_kozai}
	\mathcal{H}_{KL} = \mathcal{F}_{0} (J,J_z,J_\nu) + \mathcal{F}_{1} (J,J_z,J_\nu,\psi,\psi_z).
\end{equation}
The dependency of $\mathcal{F}_{1}$ on $\psi$ can be eliminated through averaging over this fast angle \citep{Morbidelli:1993ul,Morbidelli:2002tn}. Therefore, we get a nice and simple one d.o.f Hamiltonian of the Kozai-Lidov dynamics inside the retrograde MMR:
\begin{equation}\label{eq:H_kozai_aver}
	\overline{\mathcal{H}}_{KL} = \overline{\mathcal{F}}_{0} (\overline{J},\overline{J}_z,\overline{J}_\nu) + \overline{\mathcal{F}}_{1} (\overline{J},\overline{J}_z,\overline{J}_\nu,\overline{\psi}_z).
\end{equation}
For the sake of simplicity, from now on, we still use notations $J$, $N$ and $S_z$ to denote the averaged actions $\overline{J}$, $\overline{J}_\nu$ and $\overline{J}_z$. In addition, we replace $\overline{\psi}_z$ with $\omega$, as $\overline{\psi}_z = \overline{\sigma}_z = \overline{\sigma} + \overline{\omega}$ and for a retrograde MMR, $\sigma$ always librates around $0^\circ$ or $180^\circ$. We know for sure that the Kozai-Lidov dynamics depends on angle $2\omega$, so this replacement of angle will not alter its phase-space portrait. The final form of the Hamiltonian reads:
\begin{equation}\label{eq:H_kozai_final}
	\overline{\mathcal{H}}_{KL} = \overline{\mathcal{F}}_{0} (J,N,S_z) + \overline{\mathcal{F}}_{1} (J,N,S_z,\omega).
\end{equation}

\section{Retrograde 1:1 Resonance in 3D}\label{sec:3}

From now on, we focus on the most complicated retrograde mean motion resonance, that is, the retrograde 1:1 resonance. Unlike the prograde 1:1 resonance, where the resonant angle can librate around $0^\circ$ (quasi-satellite orbits), $180^\circ$ (horseshoe orbits), and $\pm 60^\circ$ (tadpole orbits) \citep{Giuppone:2010bp,Namouni:1999ci}, the resonant angle $\varphi$, which is equal to $(k+k^\prime)\sigma$, can only librate around $0^\circ$ (pericentric libration) and $180^\circ$ (apocentric libration) in the retrograde 1:1 resonance. The pericentric libration exists for all values of eccentricities, while the apocentric libration is absent when the eccentricity is moderate. Moreover, it is shown in \citet{Huang:2018ey} that the width of pericentric libration is much larger than that of apocentric libration. So practically, asteroids are most likely to be trapped in a resonance centered around $0^\circ$, rather than $180^\circ$. Therefore, when dealing with the Kozai-Lidov dynamics inside the 1:1 retrograde MMR, we only study the pericentric case and assume the resonant angle is librating around $0^\circ$. This is also the case of the mysterious retrograde co-orbital asteroid, 2015 BZ509 \citep{Wiegert:2017fj,Morais:2017fn}.

\subsection{Exact Resonance}

To begin with, we consider the case of exact resonance, where the resonant angle is always equal to $\sigma_0$. It can be accomplished by setting the action $J=0$ in Equation~\eqref{eq:H_kozai_final}. In practice, however, we simply set the semi-major axis $a=a_\textrm{res}$ to make the computations easier \citep{Morbidelli:2002tn} . Under this assumption, for each values of $N$, we can evaluate Equation~\eqref{eq:H_kozai_final} on a $S_z - \omega$ grid and then plot its level curves on an $e - \omega$ space. To be specific, the expression of $\overline{\mathcal{F}}_{0}$ is simply Equation~\eqref{eq:H_0}, and $\overline{\mathcal{F}}_{1}$ is now a single integral:

\begin{equation}\label{eq:averaging_single}
	\overline{\mathcal{F}}_{1} = -\cfrac{\mu}{2\pi k} \int_{0}^{2\pi k}  \left. \left(\cfrac{1}{\left| \bm{r}-\bm{r}^\prime \right|} - \cfrac{\bm{r}\cdot\bm{r}^\prime}{{\bm{r}^\prime}^3}\right) \right|_{\varphi=\varphi_0} \mathrm{d} \lambda^\prime.
\end{equation}

\begin{figure}
	\centering
	\includegraphics[width=\columnwidth]{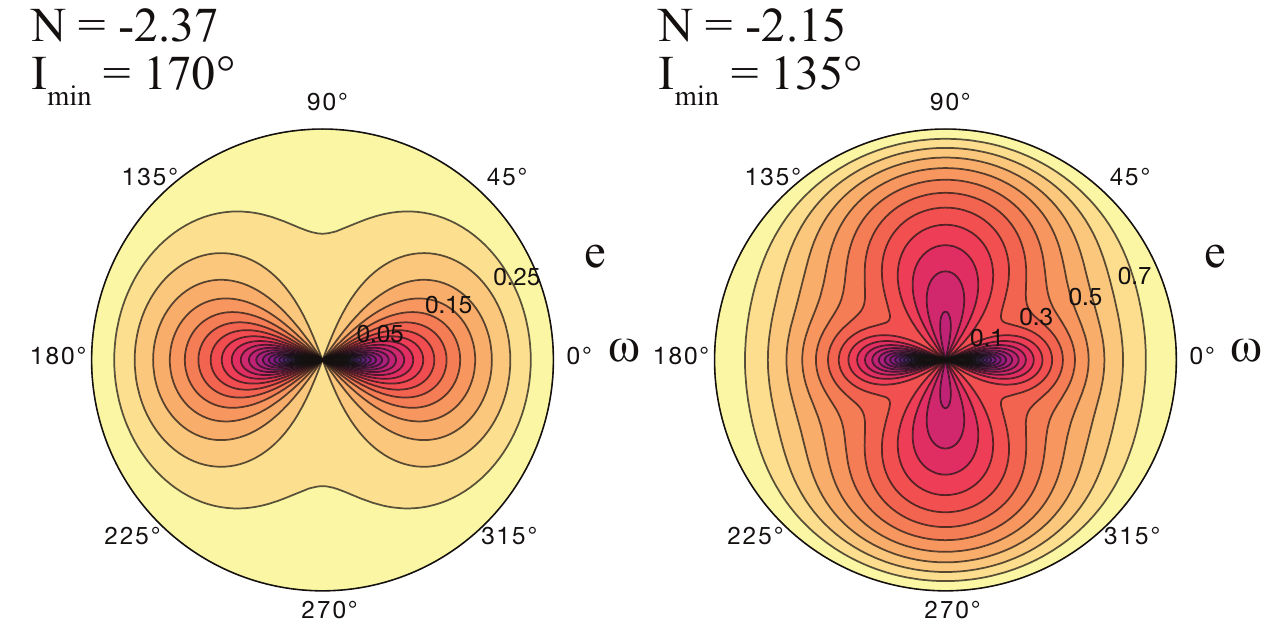}
	\caption{Typical phase-space portraits of Kozai-Lidov dynamics inside the retrograde 1:1 resonance with zero amplitude of the resonant angle $\varphi$. The level curves of Hamiltonian~\eqref{eq:H_kozai_final} are plotted on an $e - \omega$ polar space. The left panel represents the case where the eccentricity and the relative inclination (i.e. $180^\circ - I$) are moderate ($N=-1.95$, $I_{\textrm{min}}= 162^\circ$); While the right panel corresponds to the high eccentricity and relative inclination case ($N=-1.50$, $I_{\textrm{min}}= 120^\circ$).}
	\label{fig:11_ome_e_retro}
\end{figure}

We investigate different values of $N$ from $-1.99$ to $-1.30$, covering cases from small to extremely high eccentricity. But with no luck, we do not find any Kozai-Lidov librational region on the phase-space portrait. As shown in Figure~\ref{fig:11_ome_e_retro}, no matter what the value of $N$ is, there is no equilibrium point on the $e - \omega$ phase space. In other words, the argument of perihelion $\omega$ will never librate inside an exact retrograde 1:1 resonance, even though its $I_{\textrm{min}}$ \footnote{$I_{\textrm{min}}$ is equivalent to the $I_{\textrm{max}}$ in the prograde Kozai-Lidov dynamics study.} is considerably high.

However, the Figure~\ref{fig:11_ome_e_retro} does show us a tendency, that as the relative inclination increases, the Kozai-Lidov effect around $90^\circ$ and $270^\circ$ are going to strengthen. We can see it by comparing the above two phase-space portraits, and the right one has a group of clear $8$ shaped curves around $90^\circ$ and $270^\circ$. But due to the restriction of the retrograde 1:1 resonance, this tendency does not grow into two separate librational islands as the classical Kozai-Lidov dynamics does.

\subsection{The Effect of the Resonant Amplitude}

After analyzing the case of exact resonance, we reconsider the influence of the amplitude of $\varphi$ onto the Kozai-Lidov dynamics. Specifically, we enforce the resonant angle to vary along a sinusoidal curve with a given amplitude $\varphi_{\textrm{amp}}$, and then calculate the value of Hamiltonian~\eqref{eq:H_kozai_final}. The sinusoidal approximation of the evolution of the resonant angle also appears in previous works about the resonant Kozai-Lidov dynamics \citep{Gomes:2005fl,Gallardo:2012gs,Brasil:2014kr}.

Considering the imposed variation of the resonant angle $\varphi$, $\overline{\mathcal{F}}_{1}$ is now a double integral with the following form:
\begin{equation}\label{eq:averaging_double}
	\overline{\mathcal{F}}_{1} = -\cfrac{\mu}{4\pi^2 k} \oint_{S_{\varphi_{\textrm{amp}}}} \int_{0}^{2\pi k} \left(\cfrac{1}{\left| \bm{r}-\bm{r}^\prime \right|} - \cfrac{\bm{r}\cdot\bm{r}^\prime}{{\bm{r}^\prime}^3}\right) \mathrm{d} \lambda^\prime \mathrm{d} \varphi,
\end{equation}
where $S_{\varphi_{\textrm{amp}}}$ denotes a sinusoid with amplitude $\varphi_{\textrm{amp}}$.

We know the fact that when the pericentric libration occurs, the variation of $\varphi$ is not strictly a sinusoid. But this approximation makes the evaluation of the perturbation~\eqref{eq:averaging_double} much more convenient, as we do not have to integrate the equations of motion to get the time evolution of $\varphi$. And we will see later that it shows great agreement with the real Kozai-Lidov dynamics.

\begin{figure*}
	\centering
	\includegraphics[width=\textwidth]{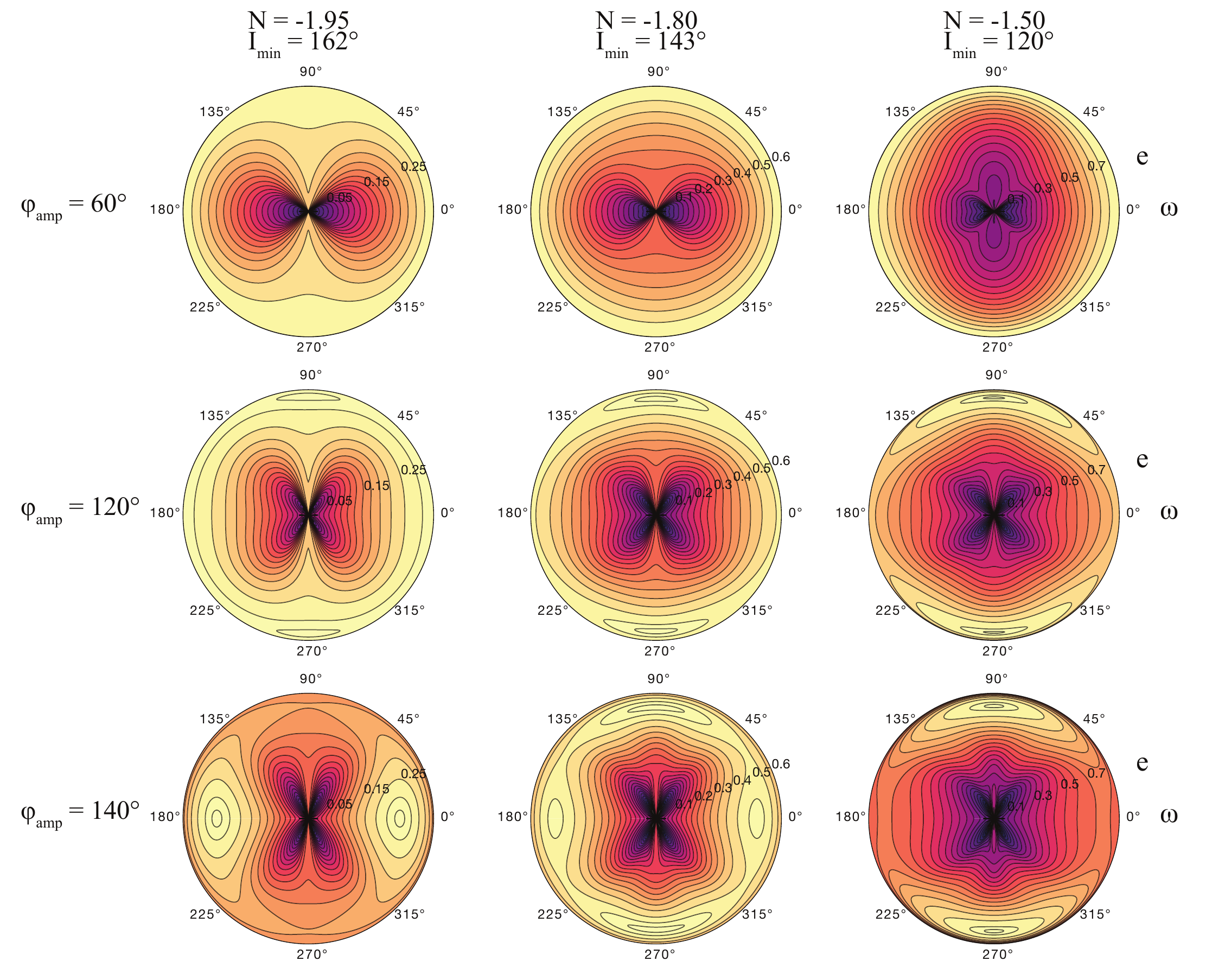}
	\caption{Phase space portraits on $e-\omega$ considering the amplitude of resonant angle $\varphi$. Three different cases of resonances ($N = -1.95, -1.80, -1.50$ and $I_\textrm{min} = 162^\circ, 143^\circ, 120^\circ$) and three values of resonant amplitude ($\varphi_\textrm{amp} = 60^\circ, 120^\circ, 140^\circ$) are taken into account. Nine graphs in total are presented above.} \label{fig:11_ome_e_comp}
\end{figure*}

As shown in Figure~\ref{fig:11_ome_e_comp}, we generate nine $e - \omega$ phase-space portraits to illustrate the effect of resonant amplitude $\varphi_\textrm{amp}$ on the Kozai-Lidov dynamics. From left to right, it represents cases from small $N$ (i.e small eccentricity and small relative inclination) to large $N$. And from top to bottom, it corresponds to $\varphi_\textrm{amp}$ from moderate to large. The largest $\varphi_\textrm{amp}$ we present here is $140^\circ$ because of the peculiarity of planar retrograde 1:1 resonance, that the pericentric region and apocentric region are separated by the collision curve around $160^\circ$ \citep{Huang:2018ey}. This means that as long as the resonant angle is near $160^\circ$, it will get strong perturbation from the planet, leading to the distortion or even disruption of the resonant dynamics.

The first row of Figure~\ref{fig:11_ome_e_comp} depicts the Kozai-Lidov dynamics when $\varphi_\textrm{amp} = 60^\circ$. We see that the level curves are not distorted too much in comparison with Figure~\ref{fig:11_ome_e_retro}, showing great resemblance to the zero-amplitude dynamics. But the dynamics becomes far more interesting as the $\varphi_\textrm{amp}$ goes to a rather larger value, such as $120^\circ$. As shown in the second row, new librational regions appear around $\pm 90^\circ$ for all of the portraits. These islands are located on the outermost edge of the graph, which indicates they originate from the near planar orbits ($I = 180^\circ$). This mechanism of new equilibrium points emerging is different from the bifurcation of the classical Kozai-Lidov dynamics, where librational islands grow from the center of the $e - \omega$ space and will not materialize until the inclination crosses a certain threshold. However, the increase of $\varphi_\textrm{amp}$ seems to provide another mechanism that prompting the libration of $\omega$ around $\pm 90^\circ$, even though the relative inclination of the orbit is small and near the plane. When the $\varphi_\textrm{amp}$ grows to an even larger value $\varphi_\textrm{amp} = 140^\circ$, things get much more complicated. We see from the third row of Figure~\ref{fig:11_ome_e_comp} that new equilibrium points emerge from $\omega = 0^\circ$ and $180^\circ$ for the first and second portraits. As for the third portrait, where eccentricity and relative inclination are high, it is clear that the actual stable islands do not appear for this value of $N$. Moreover, the stable equilibrium points change to unstable ones in the left panel, but the stability of these two points in other portraits seems not to be affected by the growth of $\varphi_\textrm{amp}$.

Besides the complicated bifurcations caused by $\varphi_\textrm{amp}$, it is worth noting that the inner region of each portrait get highly distorted as $\varphi_\textrm{amp}$ grows. There are a series of clear butterfly-shaped curves in the center of each panel, which means relatively highly inclined orbits with small eccentricity will get strong perturbation from the planet when their $\omega$ is not $k90^\circ$. Orbits in these center butterfly regions will inevitably have close encounters with the planet, inducing the disruption of the stable retrograde 1:1 resonance.

Last but no least, the region between the butterfly-shaped region and the librational region is where the $\omega$ can circulate steadily. It is like a transition between the most stable Kozai-Lidov libration and the highly disturbed circulation in the center.

Additionally, we note that our phase-space portraits are in great agreement with the numerically calculated stability map in \citet{Morais:2016kq}, in which it is shown that simultaneous librations of both $\varphi$ and $\omega$ with small amplitude (less than $50 ^\circ$) do not even exist in their Figure 8 and 9. This characteristic of the retrograde 1:1 resonance can be easily understood by our Figure~\ref{fig:11_ome_e_comp} and Figure~\ref{fig:11_ome_e_retro}, as librational islands of $\omega$ do not emerge for the exact resonance as well as the small resonant amplitude.

\subsection{Comparison with Numerical Results in CRTBP}
We have analyzed thoroughly the phase-space portraits of Kozai-Lidov dynamics inside the retrograde 1:1 resonance for different combinations of $N$ and $\varphi_\textrm{amp}$. However, some assumptions are made to get the one d.o.f approximation of this problem. So in order to check whether the above $e - \omega$ portraits represent the real Kozai-Lidov dynamics or not, we integrate several orbits in a circular restricted three body system, and then compare them with figures we get through the semi-analytical method.

\begin{figure}
	\centering
	\includegraphics[width=0.75\columnwidth]{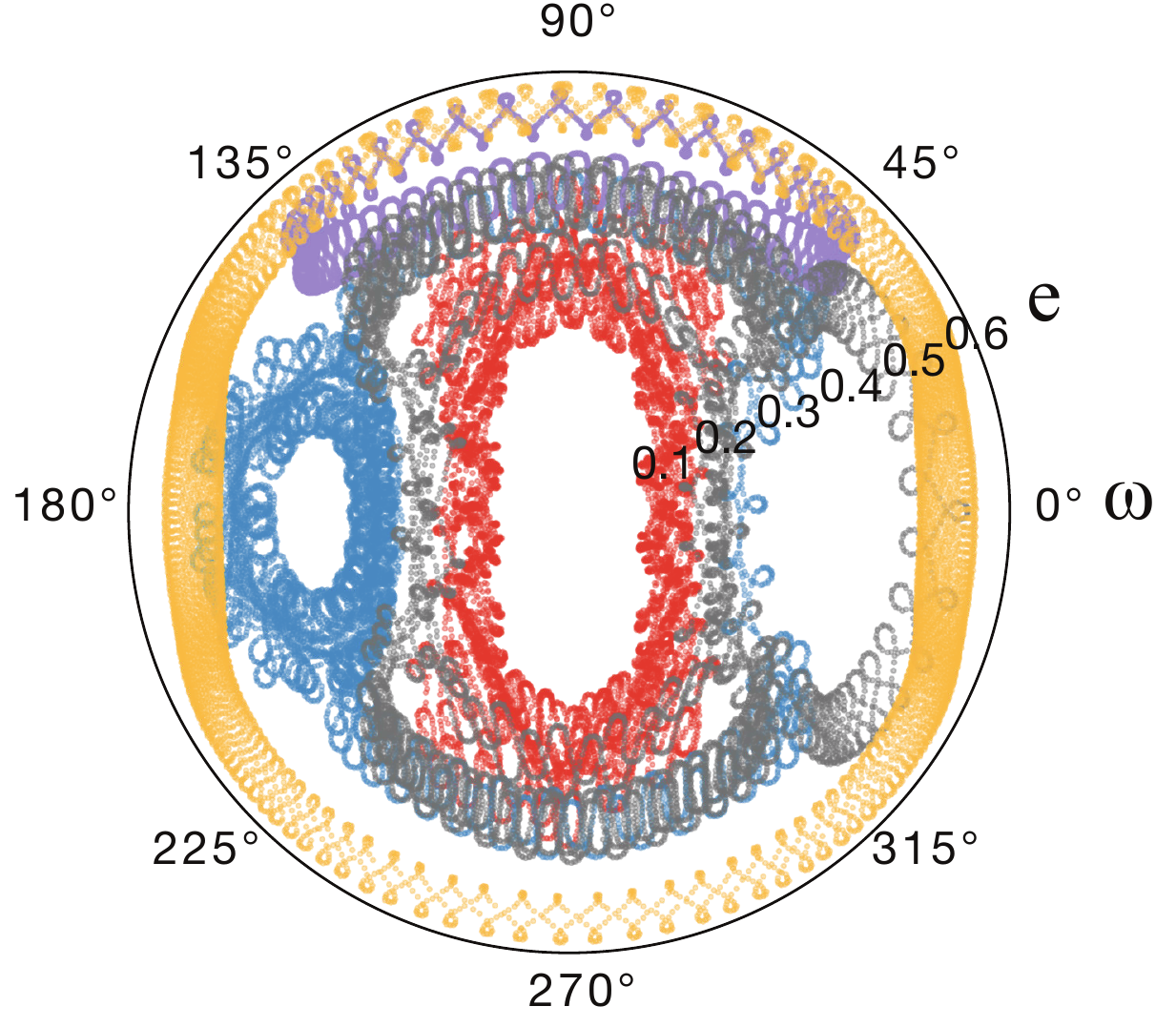}
	\caption{Numerical integration results of a group of orbits with $N = 1.8$ and $\varphi_\textrm{amp} = 140^\circ$ in a CRTBP model. Initial conditions are $a = 1$ and $I = 145^\circ, 150^\circ, 155^\circ, 160^\circ, 175^\circ$ for red, black, blue, purple, and yellow dots, respectively. Their corresponding eccentricities are calculated using the Equation~\eqref{eq:N_def}. It shows a great agreement with the middle bottom portrait in Figure~\ref{fig:11_ome_e_comp}.} \label{fig:numerical_e_w}
\end{figure}

\begin{figure}
	\centering
	\includegraphics[width=\columnwidth]{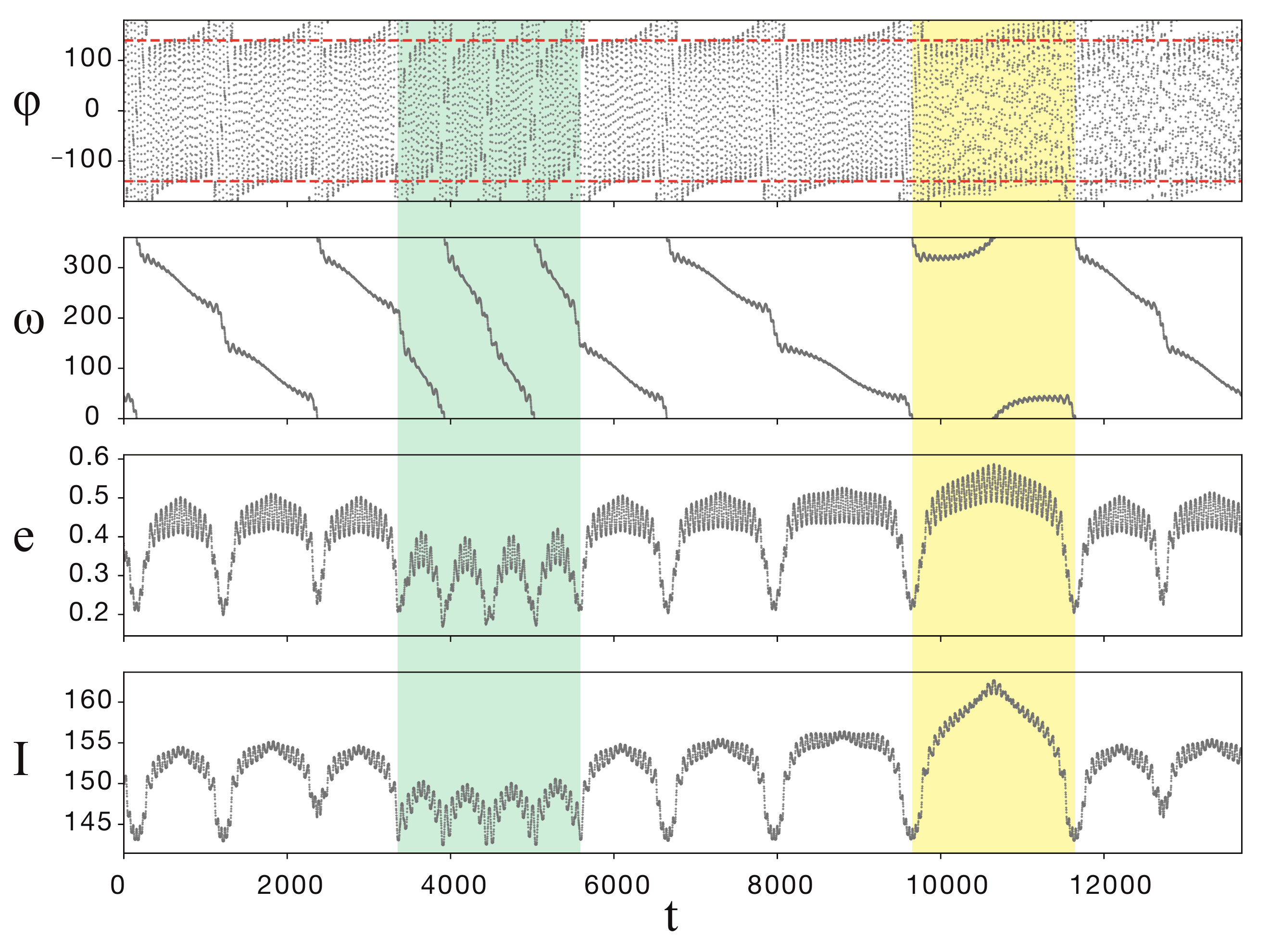}
	\caption{Dynamical evolution of the resonant angle and orbital elements for the black dots in Figure~\ref{fig:numerical_e_w}. In the first panel, $\varphi_\textrm{amp} = 140^\circ$ is presented by two red dashed lines. The green and yellow highlighted areas  are where the Kozai-Lidov dynamics alters significantly.} \label{fig:numerical_evo}
\end{figure}

As shown in Figure~\ref{fig:numerical_e_w}, the numerical results with $N = 1.8$ and $\varphi_\textrm{amp} = 140^\circ$ are in great agreement with the phase-space portrait in Figure~\ref{fig:11_ome_e_comp}. The librations around different angles (purple, blue, black) and the inner (red) and outer (yellow) circulations of $\omega$ are clearly shown in the $e - \omega$ space. Furthermore, Figure~\ref{fig:numerical_e_w} also demonstrates some divergences between the real Kozai-Lidov cycles with the smooth phase-space portrait. In particular, we focus on the blue dots and the black dots only, these orbits seem to be in a transitional situation between the circulation and the libration. Moreover, we plot the time evolution of the resonant angle and orbital elements in Figure~\ref{fig:numerical_evo} for the black dots in Figure~\ref{fig:numerical_e_w}.

We note that the real Kozai-Lidov cycles are not completely immutable as they are shown in phase-space portraits. The highlighted areas in Figure~\ref{fig:numerical_evo} indicate that the time evolution of $\omega$ may change from one circulation to another circulation (green highlighted area), or even change to a transient libration (yellow highlighted area). And it is clear that these changes occur when $\omega = 0^\circ$ or $180^\circ$. Therefore, the real evolution of $e-\omega$ does not strictly stick to the level curve in Figure~\ref{fig:11_ome_e_comp}. Instead, it is more like a stochastic wander inside a narrow area bounded by adjacent level curves in the phase space. This concept is easy to verify by comparing Figure~\ref{fig:numerical_e_w} with Figure~\ref{fig:11_ome_e_comp}. When the dot goes to the boundary between a circulation and a weak libration, in this case, where $\omega = 0^\circ$ or $180^\circ$, it may change its dynamical behaviour significantly. It is noteworthy to mention that this variation of Kozai-Lidov dynamics is not caused by planetary eccentricity or other perturber, as we carry out all the numerical calculations above in a CRTBP model.

Furthermore, we also notice that in Figure~\ref{fig:numerical_evo}, the relatively slow variation of $\omega$ actually modulates the fast angle $\varphi$. So practically, the resonant angle $\varphi$ could cross the boundary of $180^\circ$ and stay stable in the 3D case. It is clear that each time $\omega$ goes near $0^\circ$ and $180^\circ$, the resonant angle will take a transition. Therefore, not only does the resonant dynamics alter the Kozai-Lidov dynamics, but also the variation of $\omega$ has reverse impact onto its resonant dynamics.

\section{Application to Resonant Minor Bodies}\label{sec:4}

In this chapter, we are going to implement the above semi-analytical method to analyze the Kozai-Lidov dynamics of real retrograde 1:1 resonant minor bodies in the solar system. This helps us to better understand the stability of these co-orbital bodies, and may also shed light upon their mysterious origins. We pick the first confirmed asteroid in retrograde 1:1 resonance with Jupiter, 2015 BZ509, and the most potential Centaur in retrograde 1:1 resonance with Saturn, 2006 RJ2. The phase-space portraits related to these two minor bodies are depicted with their respective mean $N$ and $\varphi_\textrm{amp}$ during the integration time.

\subsection{2015 BZ509}
2015 BZ509 may be the most distinct newly found asteroid in the solar system. It is trapped in a quite stable retrograde 1:1 mean motion resonance with Jupiter, and may have an interstellar origin as pointed out by \citet{Namouni:2018hl}. We think it is more pragmatic to figure out its unique dynamical structure first, including mean-motion one and secular one, then to explore its possible root.

\begin{figure*}
	\centering
	\includegraphics[width=\textwidth]{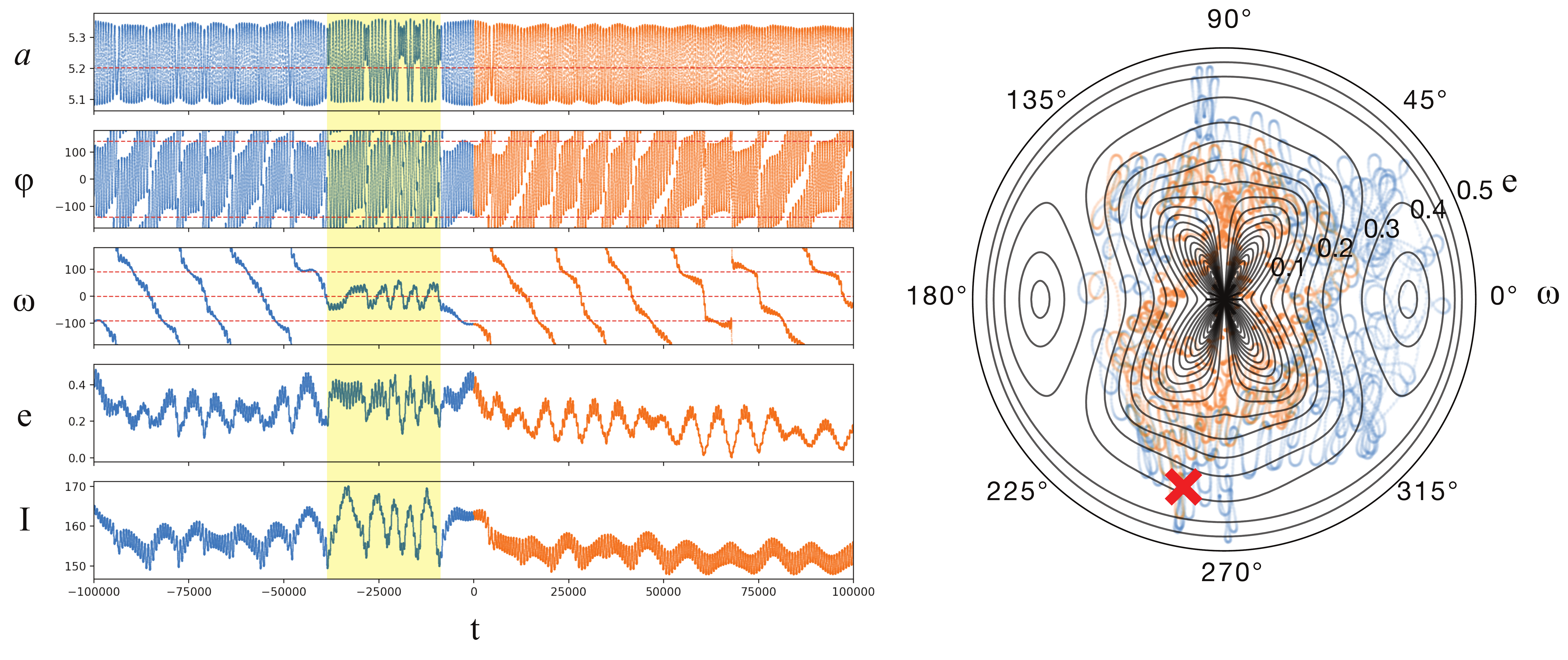}
	\caption{Left panel: dynamical evolution of orbital elements and the resonant angle of 2015 BZ509. The orbit is integrated from current epoch to 100 kyr (orange lines) and -100 kyr (blue lines) while considering all eight planets. The red dashed lines in the top three panels denotes the exact resonant value of $a$, the boundaries of $\varphi_\textrm{amp} = 140^\circ$, and $\omega = 0^\circ, \pm 90^\circ$, respectively. The yellow highlighted area is where the alternation of Kozai-Lidov dynamics occurs. Right panel: phase-space portrait generated by the semi-analytical method along with the numerical trace of 2015 BZ509 on the $e-\omega$ space. The parameters for creating this portrait are $N=-1.88$ and $\varphi_\textrm{amp} = 140^\circ$. The red bold cross denotes its current location.} \label{fig:BZ509_num_phase}
\end{figure*}

To begin with, we integrate its nominal orbit \footnote{Data are retrieved from JPL Small-Body Database Search Engine (\url{https://ssd.jpl.nasa.gov/sbdb_query.cgi}) on June 2018, same below for 2006 RJ2.} forward and backward 100 kyr, respectively, using the MERCURY \citep{1999MNRAS.304..793C} package. Then we compare the numerical integration result to its phase portrait. As shown in Figure~\ref{fig:BZ509_num_phase}, the time evolution of orbital elements and the resonant angle $\varphi$ are plotted in the left panel, while the right panel depicts its $e-\omega$ evolution and the Kozai-Lidov phase-space portrait. We note from Figure~\ref{fig:BZ509_num_phase} that during most of its integration time, the argument of pericenter $\omega$ of 2015 BZ509 is in a Kozai-Lidov circulation. However, it was captured into a Kozai-Lidov libration around $0^\circ$ about 40 kyr ago, and then left the libration recently. The phase portrait in the right panel shows regions of circulation and libration with levels curves of Hamiltonian, which corresponds to the dynamical evolution of 2015 BZ509 quite well.

However, We have to point out that due to the perturbation of other planets, and the influence caused by the eccentricity and inclination of Jupiter, the constant $N$ in the CRTBP model is not a constant any more. Similarly, the variation of resonant angle does not follow the sinusoid exactly, and the amplitude of the resonant angle is not immutable as well. Therefore, considering all the aforementioned shortages, in practice, we generate the $e-\omega$ portrait of Figure~\ref{fig:BZ509_num_phase} with the mean value of $N$ and a roughly estimated value of $\varphi_\textrm{amp}$ during the simulation. In addition, considering the modulation of $e$ and $I$ by large-amplitude mean motion resonance, the trace of numerical dots on the $e-\omega$ space is somewhat messy. The complicated coils of numerical trace shown in Figure~\ref{fig:BZ509_num_phase} are not produced by secular effect, but by mean motion dynamics.

Even though we study the dynamics with a rather simplified and approximated model, it indeed supplies us a good reference to the real Kozai-Lidov dynamics. The evolution of $\omega$ of 2015 BZ509 does change from a inner circulation to another libration centered at $0^\circ$ (yellow highlighted area), which resembles the alternation of Kozai-Lidov dynamics in Figure~\ref{fig:numerical_evo}. Besides, it is also shown in Figure~\ref{fig:BZ509_num_phase} that, during the circulation, $\omega$ may stay around $\pm 90^\circ$ for a while, when its eccentricity grows to a relatively high value. It is easy to explain this behaviour with the phase-space portrait in the right panel, that shows that the evolution of $\omega$ slows down around $\pm 90^\circ$ because of the unstable equilibrium point here. It is like a single pendulum that will stay at the top for a relatively long time when circulating.


\subsection{2006 RJ2}
In our recently published paper, \citet{Li:2018kn} identifies four candidates potentially trapped in the retrograde 1:1 resonance with Saturn, 2006 RJ2, 2006 BZ8, 2017 SV13, and 2012 YE8. Among these candidates, only 2006 RJ2 has a relatively long-time resonant state with Saturn (above 100 kyr), in which the resonant Kozai-Lidov dynamics is able to influence the orbital evolution of the minor body significantly. Therefore, in this part, we carry out the same methodology to analyze the resonant and Kozai-Lidov state of 2006 RJ2.

\begin{figure*}
	\centering
	\includegraphics[width=\textwidth]{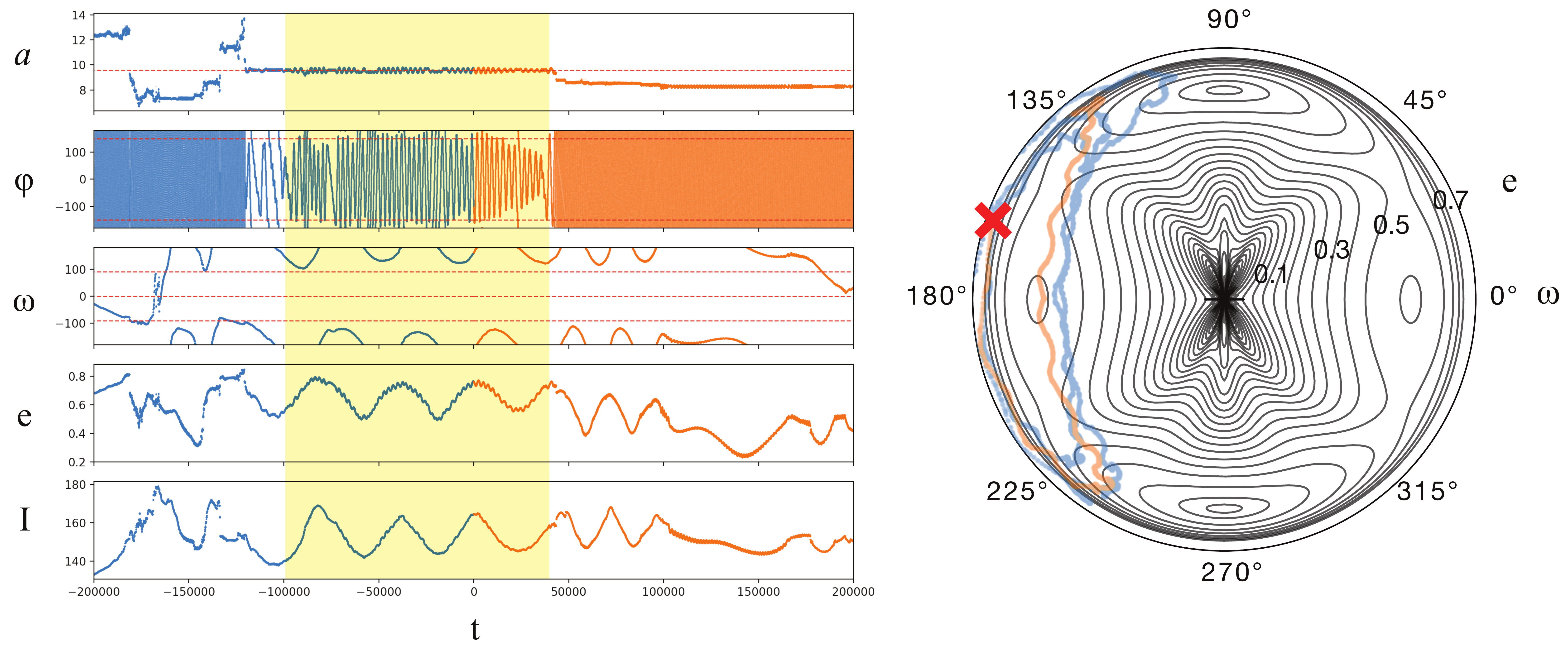}
	\caption{The basic information is the same as Figure~\ref{fig:BZ509_num_phase}, but the target asteroid is 2006 RJ2 and the integration time is extended to $\pm$200 kyr now. The right portrait is generated with the parameters $N=-1.66$ and $\varphi_\textrm{amp} = 150^\circ$. The librational segment is highlighted with yellow color in the left panel, which corresponds to the numerical trace in the right panel.}\label{fig:RJ2_num_phase}
\end{figure*}

As shown in Figure~\ref{fig:RJ2_num_phase}, among the 400 kyr integration time, 2006 RJ2 is in a good retrograde 1:1 resonance with Saturn for about 140 kyr (yellow highlighted area). Because of its relatively large eccentricity, the orbit of 2006 RJ2 will inevitably intersect with the orbit of Jupiter. Therefore, the main reason for its entering and leaving from the resonance with Saturn is that it gets two strong gravitational perturbations by close encounters with Jupiter, which is also indicated by large jumps of semi-major axis $a$ in Figure~\ref{fig:RJ2_num_phase}.

However, during its relatively stable retrograde resonance with Saturn, 2006 RJ2 has less and weaker close encounters with Jupiter, so its resonant and Kozai-Lidov dynamics are mainly shaped by Saturn at this time. Therefore, we plot the evolution of $e-\omega$ of 2006 RJ2 and its corresponding Kozai-Lidov portrait on the right panel of Figure~\ref{fig:RJ2_num_phase}, and find they match very well. The phase-space portrait embodies the librational islands around 4 different angles of $\omega$, which is caused by the large resonant amplitude $\varphi_\textrm{amp}$. And the numerical trace does librate around the $180^\circ$ center when it is inside the retrograde 1:1 resonance.

Comparing the dynamical evolution of 2006 RJ2 and 2015 BZ509, we find that the resonant states of these two minor bodies are quite different, despite their critical angles both librate around $0^\circ$ with a rather large amplitude. First of all, we see the modulation of $\varphi$ appears for 2015 BZ509 but does not emerge for 2006 RJ2. It can be explained by our phase-space portraits, that for an inner Kozai-Lidov circulation, each time $\omega$ is near $0^\circ$ and $180^\circ$, its eccentricity will drop to a comparatively small value, and the numerical trace will go near the central unstable region. On the other hand, for a stable libration around $0^\circ$ and $180^\circ$, the orbital eccentricity is always above a rather safe value, which means the minor body can not get strong perturbations from the main planet. Therefore, if we only take into account the disturbation of the main perturber, the resonant state of 2006 RJ2 should be more stable than 2015 BZ509, as the former is trapped in both the retrograde mean motion resonance and the Kozai-Lidov resonance.

However, the disturbation of Jupiter cannot be ignored in practice. The two resonances of 2006 RJ2 only prevent it from close encounters with Saturn, but do not change the fact that it will meet with Jupiter by `luck', as its eccentricity is already above the Jovian crossing threshold. Furthermore, the co-orbital region of Saturn may be intrinsically unstable due to the resonance overlap between 1:1 resonance with Saturn and 2:5 resonance with Jupiter. This mechanism is used to explain the depletion of Saturn Trojans \citep[Chapter 9]{Morbidelli:2002tn} and may also be a cause to the comparatively short life span of Saturn retrograde Trojans \citep{Li:2018kn}. Further study into this problem is beyond the scope of this paper. We will investigate the stability of the retrograde co-orbital region of Saturn and Jupiter in our future work.

Back to the asteroid 2015 BZ509, although it has a relatively longer life span of the retrograde 1:1 resonance than 2006 RJ2, it does not imply it can always retain its current resonant state. As we analyzed before, 2015 BZ509 is in a transitional state between Kozai-Lidov inner circulation and libration. Its stability is not always protected by both resonances, and its resonant angle is not in a very regular librational cycle. Additionally, the numerical trace of 2015 BZ509 occasionally passes through the inner butterfly-shaped region, where the stability is undermined by close encounters with the main perturber demonstrated by our simple CRTBP model. Therefore, by analyzing its numerical results and the corresponding phase-space portrait, we think that 2015 BZ509 is not in a very stable resonant state as thought before, and its current retrograde co-orbital resonance can get broken through close encounters with Jupiter.

\section{Other Retrograde Resonances}\label{sec:5}

In the above section, we demonstrated that the semi-analytical method is an effective tool to explore the dynamical structure of a retrograde mean motion resonance. It is in great agreement with the numerical results obtained from the CRTBP model, and also supplies a good reference to the resonant and Kozai-Lidov dynamics of the real minor bodies. On this basis, we further investigate the Kozai-Lidov cycles inside other retrograde resonances, and find totally different bifurcations of equilibrium points compared to the 1:1 resonance.

\subsection{Retrograde 2:1 Resonance}

We focus on the retrograde 2:1 mean motion resonance first. It is an inner resonance whose critical semi-major axis $a_\textrm{2/1} = 0.62996~ a_{\textrm{pla}}$. Like how we handle the retrograde 1:1 resonance, we need to understand its resonant dynamics first, using the method presented in \cite{Huang:2018ey}. We do not plot its resonant phase-space portrait here since its bifurcation is pretty simple and easy to summarize. The retrograde 2:1 resonance has two librational center, one at $\varphi = 0^\circ$ and another at $180^\circ$. Specifically, The pericentric center exists for all values of eccentricities while the apocentric center only exists when the orbital eccentricity is above the planet-crossing threshold ($e > 0.587$). And as the libration width around the pericentric center is always larger than that around the apocentric center, except for the extremely-large-eccentricity case, it is more practical to study the Kozai-Lidov mechanism inside $\varphi = 0^\circ$.

\begin{figure}
	\centering
	\includegraphics[width=\columnwidth]{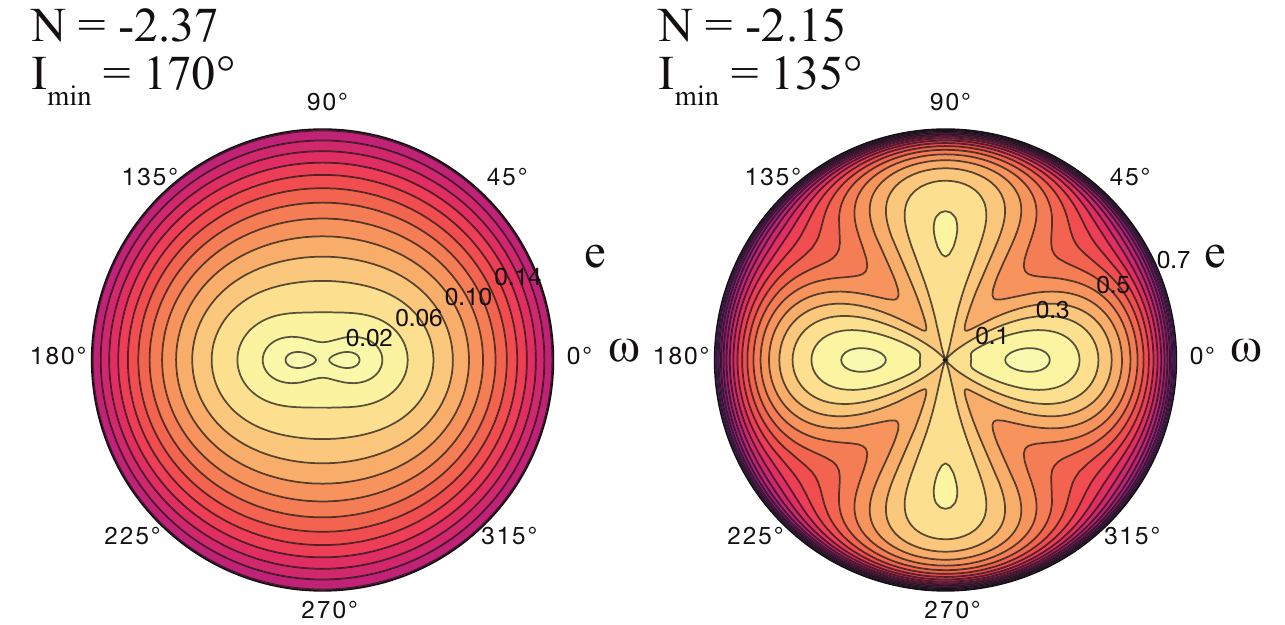}
	\caption{Typical phase-space portraits of Kozai-Lidov dynamics inside the retrograde 2:1 resonance with zero amplitude of the resonant angle $\varphi$. The left panel represents the case where the eccentricity and the relative inclination (i.e. $180^\circ - I$) are low ($N=-2.37$, $I_{\textrm{min}}= 170^\circ$); While the right panel corresponds to the high eccentricity and relative inclination case ($N=-2.15$, $I_{\textrm{min}}= 135^\circ$).}
	\label{fig:21_ome_e_retro}
\end{figure}

As shown in Figure~\ref{fig:21_ome_e_retro}, the Kozai-Lidov dynamics inside the retrograde 2:1 resonance has its unique structure. First of all, when the eccentricity is low, there are still two librational islands around $0^\circ$ and $180^\circ$. As the orbit becomes more elliptical and inclined, the two librational islands around $\pm90^\circ$ starts to emerge on the phase portrait, while the two former islands also enlarge their areas. In addition, we also try incorporating the influence of the resonant amplitude and observe that the growth of $\varphi_\textrm{amp}$ will weaken the Kozai-Lidov libration around $0^\circ$ and $180^\circ$, while has no clear effect onto the dynamics around two vertical islands. Therefore, it seems that as the resonant amplitude increases, the resonant Kozai-Lidov dynamics tends to converge with the local (i.e. same $a$,$e$, and $I$) nonresonant one. This conclusion is also true for our retrograde 1:1 case, since the nonresonant co-orbital Kozai-Lidov dynamics has all four equilibrium points around $k90^\circ$~\citep[Figure 1a-f]{Michel:1996us}.

\subsection{Retrograde 2:5 Resonance}

Another interesting retrograde resonance is the 2:5 resonance around $a_\textrm{2/5} = 1.84202~ a_{\textrm{pla}}$. This is because currently the periods of Jupiter and Saturn are near the 2:5 ratio, so the 1:1 resonance with Saturn will inevitably overlap with the outer 2:5 resonance with Jupiter. As pointed out by \citep{Li:2018kn,Morais:2013ka}, 2006 BZ8 is possibly inside a compound configuration between the retrograde 1:1 and 2:5 resonance right now. So investigation into the dynamical structure of the retrograde 2:5 resonance is essential to explain the instability of Saturnian co-orbital region.

Likewise, the libration of the retrograde 2:5 resonance occurs around $0^\circ$ and $180^\circ$. However, in this case, the apocentric libration exists for all of the eccentricities, while the pericentric libration only appears when the eccentricity crosses the planet-crossing value ($e > 0.457$). Moreover, we note that the apocentric libration width is relatively too small to have a significant impact on the asteroidal orbit when the eccentricity is lower than $0.4$. As a result, we may ignore the low-eccentricity case and consider only the highly eccentric orbit.

\begin{figure}
	\centering
	\includegraphics[width=\columnwidth]{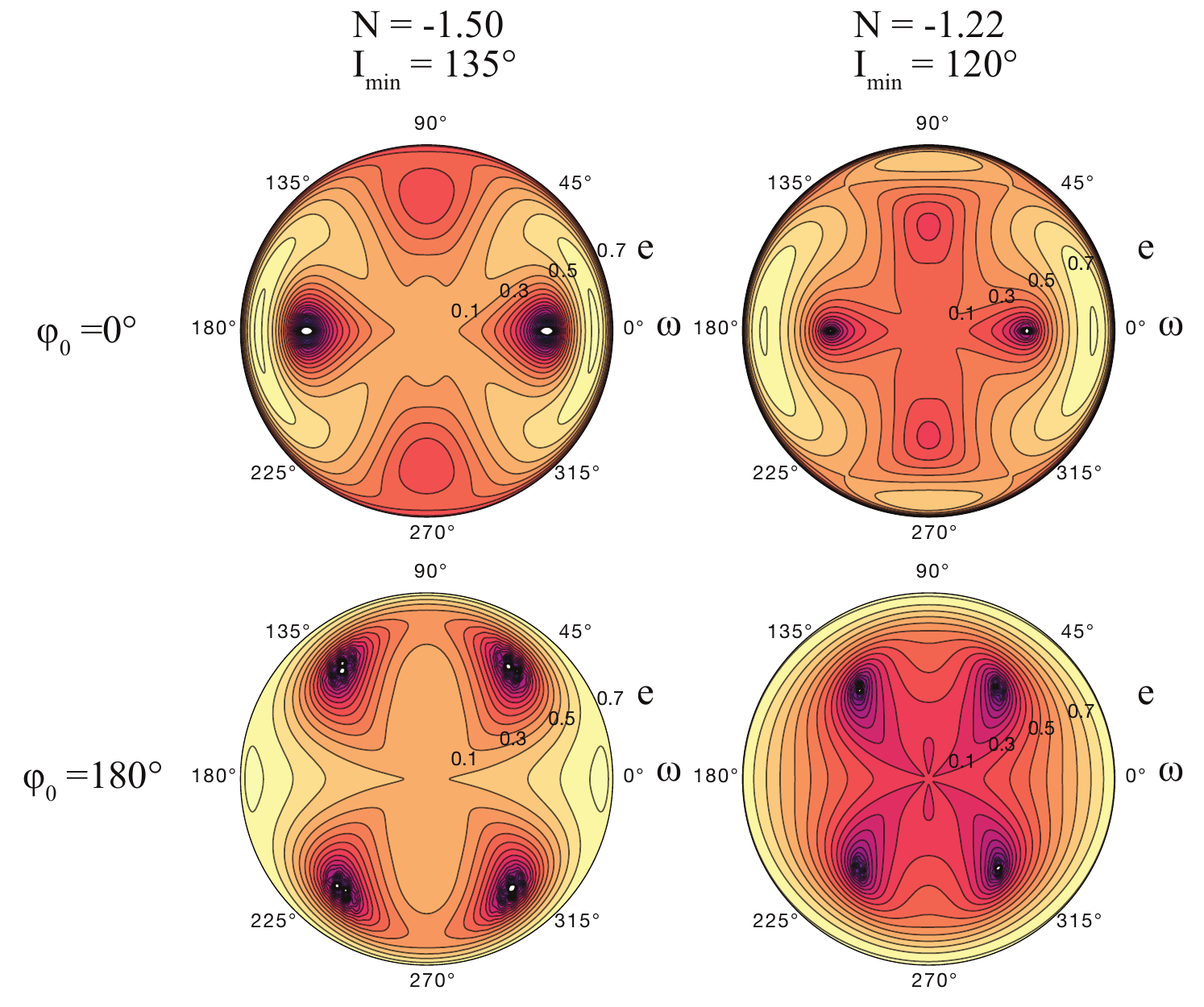}
	\caption{Typical phase-space portraits of Kozai-Lidov dynamics inside the retrograde 2:5 resonance with zero amplitude of the resonant angle $\varphi$. The parameters used to generated these four figures are $N=-1.50$ and $I_\textrm{min}=135^\circ$ for the left panel, and $N=-1.22$ and $I_\textrm{min}=120^\circ$ for the right panel. And the resonant center of the top panel is $0^\circ$ while that of the bottom panel is $180^\circ$. The black dense circles denote where close encounters with planet occur, and evolutions around these regions seem to be unstable.}
	\label{fig:25_ome_e_retro}
\end{figure}

Figure~\ref{fig:25_ome_e_retro} presents four phase-space portraits whose librational centers are $0^\circ$ and $180^\circ$, and minimum inclinations are $135^\circ$ and $120^\circ$, respectively. It is clear that the shift of the librational center has significant impact on its Kozai-Lidov dynamics. Specifically, for the pericentric libration in the top left panel, there are four librational islands around $k90^\circ$, and two unstable collisional regions (black dense circles on Figure~\ref{fig:25_ome_e_retro}) around $\omega = 0^\circ$ and $180^\circ$. However, for the apocentric libration, the unstable regions split to four and occupy the area around $\omega = k45^\circ$, while the librational islands only appear at $0^\circ$. As the eccentricity grows, another librational islands emerge around $\pm90^\circ$ for the pericentric case. On the other hand, the two librational islands vanish for the apocentric case, which are shown in the bottom panel of Figure~\ref{fig:25_ome_e_retro}.

We note that from Figure~\ref{fig:11_ome_e_retro}, Figure~\ref{fig:21_ome_e_retro}, and Figure~\ref{fig:25_ome_e_retro} that the bifurcation of Kozai-Lidov dynamics for each retrograde resonance is notably different. Sometimes the exact resonance inhibits the occurrence of the Kozai-Lidov cycles, whereas for other ratios, the resonant dynamics compels the argument of pericenter librating around four critical angles. Therefore, it seems to be tough to extract an overall conclusion on how the retrograde resonance alters the local Kozai-Lidov dynamics. But we do present an effective approach to analyze a particular retrograde resonance thoroughly, and it can be applied to resonances with arbitrary ratios.

\section*{Conclusion}

In this paper, we present a complete and self-contained methodology to analyze the Kozai-Lidov dynamics inside the retrograde mean motion resonance. It is an extension work of our last paper \citep{Huang:2018ey}, in which the planar dynamics of a retrograde resonance can be well understood through the semi-analytical method. Combined with this work, we can now handle the retrograde resonance in 3D space.

We firstly deduce the canonical transformations that can be applied to retrograde orbits, and obtain a nice and simple one d.o.f integrable approximation for this particular problem. With the dominant Hamiltonian in hand, it is easy to carry out a numerical averaging process over the fast angle in closed form, and plot the level curves of Hamiltonian on a $e-\omega$ space, which are exactly the phase-space portraits we need to understand the Kozai-Lidov mechanism.

Then, we reanalyze the dynamics of the retrograde 1:1 resonance in 3D space. It is worth mentioning that in the exact resonance and small resonant amplitude case, the libration island of $\omega$ does not exist. The equilibrium points only emerge when the amplitude of the resonant angle is large enough (around $120^\circ$). Subsequently, we demonstrate with the numerical integration results, that these phase-space portraits corresponds very well to the actual Kozai-Lidov cycles in a CRTBP model. But it is also shown that the Kozai-Lidov cycle of a massless particle may change to adjacent state, even in such a simple model.

Furthermore, we apply this method to analyze the Kozai-Lidov state of real minor bodies, 2015 BZ509 and 2006 RJ2. We note that our method supplies a good reference to their Kozai-Lidov cycles, and also gives us a certain degree of information about the stability of these minor bodies. With the analysis, we think that 2006 RJ2 is currently in a relatively stable retrograde resonance with Saturn, but its stability can be easily undermined by close encounters with Jupiter due to the high orbital eccentricity. On the other hand, the resonant state of 2015 BZ509 is not as good as 2006 RJ2, as it is now positioned at a transitional state between the Kozai-Lidov libration and inner circulation. It means that its resonant state with Jupiter can be altered.

Last but not least, we further investigate the phase-space structure of the Kozai-Lidov dynamics inside other two resonances, the 2:1 and the 2:5. With the same workflow, however, we find different dynamical bifurcations on the phase-space portraits. And it turns out to be difficult to draw a simple conclusion about how the Kozai-Lidov dynamics is changed by the retrograde resonance. Maybe the better way is to analyze the interesting resonance specifically with this methodology. The only conclusion we can make here is that as the resonant amplitude grows from $0^\circ$ to $180^\circ$, the resonant Kozai-Lidov dynamics tends to get closer to its local nonresonant one.

\section*{Acknowledgements}

This work is supported by the National Natural Science Foundation of China (Grant No.11772167).




\bibliographystyle{mnras}
\bibliography{ref_2} 


\bsp
\label{lastpage}
\end{document}